\documentclass[a4paper,12pt]{article}

\usepackage{amsmath,amssymb,amsfonts}
\usepackage{graphicx}
\usepackage{epsfig}
\usepackage{setspace}
\usepackage{color}
\usepackage{amsmath}
\usepackage{amsfonts}
\usepackage{verbatim}
\usepackage{amssymb}  
\usepackage{graphicx}
\usepackage{amsmath,amssymb,calc}
\usepackage{setspace}
\usepackage{color}
\usepackage{amsmath}
\usepackage{amsfonts}
\usepackage{verbatim}
\usepackage{amssymb}
\usepackage{array}
\usepackage{caption}
\usepackage{subcaption}
\usepackage{epstopdf}
\usepackage[pdf]{pstricks}
\usepackage{upgreek}
\usepackage{cmll}

\usepackage{latexsym}
\usepackage{braket}

\usepackage{cite}
\usepackage{amsmath,amssymb,amsfonts,graphicx}
\usepackage{epsfig}
\usepackage[left=2.4cm,top=3.3cm,right=2.4cm,bottom=3.3cm,bindingoffset=0cm]{geometry}

\newcommand{\beq}{\begin{eqnarray}}
\newcommand{\eeq}{\end{eqnarray}}
\newcommand{\bea}{\begin{eqnarray}}
\newcommand{\eea}{\end{eqnarray}}
\newcommand{\be}{\begin{equation}}
\newcommand{\ee}{\end{equation}}

\def\1{\mathbbm{1}}

\def\trho{\widetilde{\rho}}
\def\tr{{\widetilde r}}
\def\tt{{\widetilde t}}
\def\tz{{\widetilde z}}

\numberwithin{equation}{section}

\begin{document}

\title{
\begin{flushright}\ \vskip -1.5cm {\small {IFUP-TH-2018}}\end{flushright}
\vskip 0pt
\bf{ \Large  On Some Universal Features of the Holographic Quantum Complexity  of Bulk Singularities}
\vskip 10pt}
\author{{\large  Stefano Bolognesi$^{(1)}$, Eliezer Rabinovici$^{(2,3)}$  and Shubho R. Roy$^{(4)}$} \\[20pt]
{\em \normalsize
$^{(1)}$Department of Physics ``E. Fermi'', University of Pisa
  and INFN Sezione di Pisa}\\[0pt]
{\em \normalsize
  Largo Pontecorvo, 3, Ed. C, 56127 Pisa, Italy}\\[3pt]
{\em \normalsize
$^{(2)}$Racah Institute of Physics, The Hebrew University of Jerusalem, 91904, Israel
}\\[3pt]{\em \normalsize
$^{(3)}$Institut des Hautes Etudes Scientifiques, 35, route de Chartres, }\\{\em \normalsize91440 Bures-sur-Yvette, France
}\\[3pt]{\em \normalsize
$^{(4)}$Department of Physics, Indian Institute of Technology, Hyderabad}\\
{\em \normalsize Kandi,
Sangareddy 502285, Medak, Telengana, India
}\\[10pt]
{\normalsize emails: stefano.bolognesi@unipi.it, eliezer@vms.huji.ac.il, sroy@iith.ac.in  } \\
{\normalsize 
}}
\vskip 0pt
\date{February 2018}
\maketitle
\vskip 0pt

\begin{abstract}

We perform a comparative study of the time dependence of the holographic quantum complexity of some space like singular bulk gravitational backgrounds.
This is done by considering the two available notions of complexity, one that relates it to the maximal spatial volume and the other that relates it to the classical action of the Wheeler-de Witt patch. We calculate and compare the leading and the next to leading terms and find some universal features. The complexity decreases towards the singularity for both definitions, for all types of singularities studied.
In addition the leading terms have the same quantitative behavior for both definitions in restricted number of cases and the behaviour itself is different for different singular backgrounds.
The quantitative details of the next to leading terms, such as their specific form of time dependence, are found not to be universal.  They vary between the different cases and between the different bulk definitions of complexity. We also address some technical points inherent to the calculation.

\end{abstract}

\newpage


\section{Introduction}

Spacetimes obtained as solutions to classical general relativity contain many types of singularities - timelike, null and spacelike. Some lurk behind event horizons while others such as bangs and crunches present themselves upfront.  Recently singularities, called firewalls, were claimed to 
emerge in places, for example near black hole horizons, where they seem absent within the classical approximation. One may expect that the deadlier ones will be healed in a quantum theory of gravity such as string theory and indeed some time like singularities have been resolved.  One is even willing  to entertain the idea that in a quantum theory one may coexist with some classes of space like singularities \cite{Barbon:2010gn,Maldacena:2010un,BR,Barbon:2013nta}. 
The framework of AdS/CFT and the accompanying bag of tools from QFT has helped make the study of some of these cases more precise.
In the context of black hole physics this enabled to test some aspects of the firewall suggestion. It was noted on the bulk side that a new very large time scale emerged, one which is associated with the growth of the volume of the Einstein-Rosen bridge in the interior of an eternal black-hole in AdS.  It grows mostly linearly in time for a period $\exp(S)$, where $S$ is the entropy of the system. This among other considerations led to the conjecture that it is related 
on the boundary Quantum Field Theory to a concept borrowed from quantum information theory-Quantum Complexity. 

This recently added item to the dictionary of the holographic correspondence between bulk and boundary is far from being defined sharply enough in either the Quantum Field Theory or the bulk system. 
 Complexity of a quantum state, somewhat loosely, is a measure of how difficult it is to obtain that state starting from a reference state and acting on it with a small number of quantum operators belonging to a  pre-defined set. 
Up to now there exists neither a  universal definition of complexity nor a complete study of its possible universality classes.
Nevertheless some universal properties are believed to be satisfied by this notion.
In particular complexity is  generically expected to continue to grow long after the maximum entropy $S$ of the system is reached at a time which is of polynomial order in $S$.  It is suggested \cite{BrownSusskind,volume}
 by some to have a long period of linear growth until it saturates at a maximum value of the order of $\exp(S)$, this by a time which itself is also of the order of $\exp(S)$. 
We remark that the claim for the existence of a maximum value of the complexity resides in the attempt to reconcile the 
continuous structure of the space of all quantum states with the manner by which each such state is reached by a discrete and finite number of prescribed operations. This is done by assigning an arbitrary required accuracy for declaring a particular approximation of the target state as successful. Without such a cutoff the complexity defined in this way has no upper bound. 
Some authors, \cite{BrownSusskind}  and references therein, have been flirting on and off with considering metrics and geodesics on the space of states  with which the system would have a universal maximum complexity, however at this stage neither this nor a dynamical cutoff were identified. The need for an arbitrary cutoff has not been removed and the states of a Quantum Field Theory have indeed a maximum complexity for given values of the entropy and the cutoff.  For other recent developments on the definition of complexity in QFT see \cite{Jefferson:2017sdb,Chapman:2017rqy,Caputa:2017yrh,Moosa:2017yiz}.

There are presently two suggestions, each with its own motivation,  as to what the complexity of a quantum boundary state should be given by in the bulk. One is that is should be, up to some proportionality coefficient,  the volume of the maximal spatial surface extending into the bulk and terminating on the boundary at the spacial slice on which the boundary quantum state is defined \cite{volume}. This is referred as the complexity proportional to volume (${\cal C}  \propto {\cal V}$) conjecture.

The other is  \cite{Brown:2015bva,Brown:2015lvg}  is that the complexity is proportional to the classical value of the action in a WdW patch. The so called 
 (${\cal C}  \propto {\cal A}$) conjecture. 
One takes the spatial slice on the boundary on which the state is defined. Then one considers the union of all possible spatial slices which extend into the bulk and terminates on the same spatial slice at the boundary. 
The union forms a sub-set of the bulk called the Wheeler-de Witt  (WdW) patch.  The Einstein-Hilbert (EH) action evaluated on this patch, with the inclusion of the proper York-Gibbons-Hawking (YGH) boundary term, is conjectured to be proportional to the complexity.  The spatial slice of maximal volume considered for the  ${\cal C}  \propto {\cal V}$  conjecture is contained in the WdW patch but now it does not play any preferential role with respect to the others.  
Be the complexity of a state what it may be on the boundary CFT  and be the corresponding bulk quantity what it may be for the purpose of the study of
singularities it is one of its features, its time dependence, that is proposed as a diagnostic tool on its own. It is on the universal features of this
quantity that we focus on in this paper.

Based on examples \cite{Brown:2015bva,Brown:2015lvg}  it is argued that as long as the complexity of a state increases in a black hole context no singularity is encountered.  In addition it is claimed that this result is robust  and remains valid when small perturbations are added to fine tuned states such as the thermofield double \cite{Stanford:2014jda}. On the other hand if the complexity does decrease in time the state may well encounter a singularity that one could associate with the formation of a firewall like obstacle. However this evolution is claimed, on the basis of examples, not to be robust, once small perturbations are added to such a system the potential formation of a firewall type singularity is delayed for at least a time of the order $\exp(S)$.
There are other circumstances in which the complexity is expected to decrease \cite{Susskind:2015toa}.
The Poincare recurrences which occur in QFTs, when they obey certain conditions, after the passage very large times of order $\exp(\exp(S))$, bring down the complexity with them. This decrease need not signal necessarily an approach to a singular configuration.  This decay is indeed not robust itself but
once one decides on what the Hamiltonian and the state are these decays will occur  again and again with Poincare time intervals.

In addition to the black hole case it was found, using the   ${\cal C}  \propto {\cal V}$  conjecture,  that the complexity does decrease as the system approaches some pre-engineered classes of time dependent singularities \cite{Barbon:2015ria}. These are intriguing singularities of the type one may be able to coexist with. 
This result could be expected and is robust for BKL \cite{Belinski:1973zz,Belinsky:1982pk}
type singularities but the generality of this feature is less clear and was obtained using the extremal volume prescription to calculate the complexity. 

Given the scarcity of the present knowledge on the nature of these singularities we explore here if this feature depends on the well motivated, but particular, bulk quantity which was conjectured to be associated with the QFT complexity. In this work we obtain the complexity by using the ${\cal C}  \propto {\cal A}$ prescription. This way of  evaluating the complexity presents its own challenges. The work will be largely of a technical nature.
The result that we find is qualitatively similar the one obtained with the   ${\cal C}  \propto {\cal V}$, in particular complexity always decreases toward the crunch. The quantitative details and in particular the specific form of time-dependence of the complexity are instead different.

Complexity is a UV divergent quantity so its regularization is necessary, as we discussed it is after the time dependence of the complexity that we are after. Both  the ${\cal C}  \propto {\cal V}$   and ${\cal C}  \propto {\cal A}$  conjectures are sensitive to this divergence and a way to deal with it is to introduce a UV cutoff close to the boundary. 
For the  ${\cal C}  \propto {\cal A}$  there is  computational difficulty. The boundary of the WdW patch is a light-like sub-manifold with joints. Both issues, being light-like and having joints, means that the YGH boundary term must be properly defined. This problem has been recently discussed in \cite{Lehner:2016vdi,Reynolds:2016rvl,Carmi:2016wjl}. For the present paper we take a different approach to this problem (see also \cite{Brown:2015lvg,Parattu:2015gga}). Since we have anyhow to UV regularize the WdW patch, we use this opportunity to introduce a particular regularization that makes the WdW boundary time-like and also smooths-out the joints. In this way we can compute the YGH term with no ambiguities.

There are two distinct types of time-dependent backgrounds. For the first type time dependence is explicit in the UV part of the metric. This case corresponds to some time dependent marginal operator in the dual field theory. The Kasner metric and the topological crunch are two examples of this type and  we will study them in detail.
In those cases also   in the UV divergent part of the complexity is time-dependent. 
For the second class the UV metric has no explicit time dependence but the whole bulk metric does. 
The de Sitter crunch model is the specific example of this type that we will consider here. 
In those last cases  the divergent piece of the complexity is free from any time dependence and the time derivative of the complexity is a finite quantity.

The paper is organized as follows.  In Section \ref{due}  we consider the cases of time dependence on the boundary. In Section \ref{tre} we consider the bulk time dependence. We conclude in Section \ref{quattro}.

\section{Time dependence in the boundary metric:\\    Kasner and Topological crunch}
\label{due}

In this section we perform a comparative calculation and study of the complexity and its time dependence as obtained according to two of its different bulk definitions. This for two cases engineered in such a way that the world volume of the boundary theory itself is singular. 
For each definition and each class of metrics we obtain the leading and next to leading terms. The result will coincide with those of  \cite{Barbon:2015ria} for the leading term when the volume prescription is adopted, this for both types of metrics.

We do this by studying two cases in which the metric on the world-volume of the dual field theory has  the  form
\beq
\label{gmet}
ds^2 = \frac{L^2}{z^2} \left(- dt^2 + dz^2 + h_{ij}(t,x) dx_i dx_j  \right) 
\eeq
with $i,j= 1, \dots, d$. 
The UV boundary is set at $z=0$ and the world-volume frame where the CFT is realized is $ds^2_{\rm CFT} = - dt^2 + h_{ij}(t,x) dx_i dx_j $. 
All cases we consider satisfy everywhere
\beq
\label{cosmo}
R -2 \Lambda = -\frac{2(d+1)}{L^2}
\eeq
where $\Lambda =-\frac{d(d+1)}{2L^2 } $ is the cosmological constant. This is true whenever the world-volume frame is Ricci flat.

The first specific example is that of the Kasner metric.
The Kasner metric  is  of the form (\ref{gmet}) with 
\beq
h_{ij}(t,x) = {\rm diag} \left( \left( {\tiny \frac tl} \right)^{2 p_1}, \dots , \left( {\tiny \frac tl} \right)^{2 p_d}  \right)
\eeq
with the relation
\beq
\label{relkas}
\sum_i p_i = \sum_i p_i^2 =1 \ .
\eeq
and $l$ is a dimensional scale.
The second specific example  is that of the so called topological crunch. 
The topological crunch,  close to the boundary, is of the form  (\ref{gmet})  with 
\beq
\label{hmettop}
h_{ij}(t,x) dx_i dx_j = l^2 \left( d \Omega_{d-1}^2 + \cos^2{\left( {\tiny \frac tl} \right)} d\phi^2 \right) \ .
\eeq
Both the Kasner metric and that of the topological crunch satify (\ref{cosmo}). 
The crunch happens when the CFT world-volume goes to zero, and this is $t=0$ for Kasner and $t=\pm  l \frac{\pi}{2} $ for the topological crunch case.  The latter is called topological because locally the metric (\ref{gmet}) with (\ref{hmettop}) is always the same as AdS$_{d+2}$.
We will perform the computations using the generic form (\ref{gmet}) and then adapt to the specific cases at the end.

Complexity, in any of its forms,  has always a divergent term that is due to the contribution coming from close to the UV boundary. 
This can be regularized by introducing a UV cutoff $\Lambda_{\rm UV}$ related to a minimal value of the $z$ coordinate
\beq
\Lambda_{\rm UV} = \frac{1}{z_{\rm UV}} \ .
\eeq

The first definition of complexity is  ${\cal C} \propto  {\cal V}$ where ${\cal V}$ is the  spatial slice with maximal volume whose boundary is at a fixed time $t= t^*$.  We consider any possible spatial slice defined by
\beq
t = f (z,x_i)
\eeq 
with the property
\beq
\lim_{z \to 0 } f(z,x_i) = t^* , \qquad \forall x_i \ .
\eeq
Since the spaces we are considering are homogeneous in $x_i$, we can always restrict to the case $\partial_i f =0$ everywhere, so from now on we take $f(z)$ to be just a function of $z$. 
The induced metric on this space slice is
\beq
ds^2 = \frac{L^2}{z^2} \left((1-(\partial_zf)^2) dz^2 + h_{ij}(f,x) dx_i dx_j  \right) 
\eeq
and the space volume is then
\bea
\label{vol}
{\cal V} &=&  \int dz  dV_x \, \, \frac{L^{d+1}}{z^{d+1}}  \sqrt{ (1-(\partial_zf)^2) \  h(f(z),x) }
\eea
where we indicate
\beq
dV_x \, = dx_1\dots dx_d \qquad {\rm and} \qquad
h(t,x) = {\rm det}\,   h_{ij}(t,x) \ .
\eeq
The expansion of the metric determinant is 
\beq
h(t,x) \simeq  h(t^*,x) +  (t-t^*) h(t^*,x) h^{ij}(t^*,x) \partial_t h_{ij}(t^*,x) + \dots 
\eeq
Note that for spaces homogeneous in $x$ the quantity  $h^{ij}(t,x) \partial_t h_{ij}(t,x)$ does not depend on $x$ so we will denote it simply as $h^{ij}\partial_t h_{ij}(t)$.

The Euler-Lagrange equation for $f(z)$, after some simplification, becomes
\beq
\label{eqf}
 \partial_z^2f  = (1-(\partial_zf)^2) \left( \frac{(1+d)\partial_zf}{z} - \frac{1}{2}h^{ij} \partial_t h_{ij}(f(z)) \right)
\eeq
By expanding near $z=0$ and solving the equation in a power series of $z$ we find that $f(z)$ goes to $t^*$ with zero derivative 
\beq
\label{expf}
f(z) =  t^*  + \alpha(t^*) z^2 +\dots \ .
\eeq
The space volume is  dominated by the following UV divergent term:
\bea
\label{vol}
{\cal V} &\simeq& \int_{1/\Lambda_{\rm UV}}  dz  \int dV_x \, \frac{L^{d+1}}{z^{d+1}}  \sqrt{  h(t^*,x) } \nonumber \\
&\simeq&   \frac{\Lambda_{\rm UV}^{d}L^{d+1}}{d}    \int dV_x \,   \sqrt{  h(t^*,x) }  \ .
\eea
Note that the leading divergence is proportional to $\Lambda_{\rm UV}^{d}$ and does not depend on  $\alpha(t^*) $ or any of the sub-leading terms in the expansion of (\ref{expf}). Note also that this term is proportional to the boundary space volume  $
 \int dV_x \, \sqrt{  h(t^*,x) }  $ and this is what provides  the dependence with respect to anchored time $t^*$.

We next turn to the sub-leading divergencies which are affected by the coefficient $\alpha(t^*) $ in (\ref{expf}).
Continuing the power expansion of the equation (\ref{eqf}) it can be found, for $d>1$ 
\beq
 \alpha(t^*) = \frac{h^{ij} \partial_t h_{ij}(t^*)}{4 d } \ .
\eeq
Note that $ \alpha(t^*)$ is uniquely determined once the first constant of integration $t^*$ is decided.
The second constant of integration appears at orders higher than $z^{2}$ in the power expansion for $f(z)$.
The sub-leading divergent term of the volume depends only on $ \alpha(t^*) $ and is 
\bea
\label{volsub}
{\cal V} =  {\cal O}( \Lambda_{\rm UV}^d) +  \Lambda_{\rm UV}^{d-2} \frac{L^{d+1} (d-1) (h^{ij} \partial_t h_{ij}(t^*))^2}{8 d^2 (d-2)} \int dV_x \,   \sqrt{  h(t^*,x) } + \dots
\eea
where ${\cal O}( \Lambda_{\rm UV}^d)$ is the leading term (\ref{vol}) while the sub-leading term is of order ${\cal O}( \Lambda_{\rm UV}^{d-2})$. Sub-leading divergent terms are present whenever $h^{ij} \partial_t h_{ij} \neq 0$.

For the case of the Kasner metric we have 
\beq
\label{kasv}
\int dV_x \, \sqrt{  h(t^*,x)}   =\frac{V t}{l}
\eeq
where $V$ is an IR cutoff for the spatial volume in $x_i$. 
Moreover for the  Kasner metric we have $
h^{ij} \partial_t h_{ij} = \frac{2}{t}
$.
The volume  is then
\beq
\label{volsubkas}
{\cal V} = \Lambda_{\rm UV}^{d} \frac{L^{d+1} V t}{d l }  + \Lambda_{\rm UV}^{d-2} \frac{ L^{d+1} (d-1)V}{2 d^2 (d-2)  l t }   + \dots
\eeq
For the leading divergence we recover the result of \cite{Barbon:2015ria}.
Note that the leading order divergent term is linear in $t$ and it is decreasing in time till it would have vanished at the singularity $t \to 0$. The validity of the approximation breaks down as one reaches to Planckian and string scale
distances $t_{\rm UV} \simeq \frac{1}{ \Lambda_{\rm UV}}$. 
This is also the time by which the sub-leading divergent term becomes of the same order of the leading one. The complexity decreases at a rate linear in time as long as the approximation is valid. 
A linear dependence of the complexity on time was obtained in the eternal black hole case among other ones \cite{volume,Brown:2015bva}, although in the latter case it was for a non-divergent piece of the complexity. 
In the Kasner  case the linear dependence follows directly from the relation (\ref{relkas}) and is not a universal feature as we are going to see next.

For the specific case of the topological crunch we have 
\beq
\label{topv}
\int dV_x \, \sqrt{  h(t^*,x)}   =  2 \pi l^d V_{S^{d}}  \cos{\left( {\tiny \frac tl} \right)} 
\eeq
and
$h^{ij} \partial_t h_{ij} = -\frac{2}{l} \tan{\left( {\tiny \frac tl} \right)}$. 
The volume is then
\bea
\label{volsubtop}
{\cal V} = \Lambda_{\rm UV}^{d} \frac{L^{d+1}  2 \pi l^d V_{S^{d}} \cos{\left( {\tiny \frac tl} \right)}   }{d} +   \Lambda_{\rm UV}^{d-2} \frac{L^{d+1}    \pi (d-1)  l^d V_{S^{d}} \sin^2{\left( {\tiny \frac tl} \right)} }{  d^2 (d-2)  l^2  \cos{\left( {\tiny \frac tl} \right)} }  + \dots
\eea
For the leading divergence we again recover the result of \cite{Barbon:2015ria}.
As in the Kasner metric case the complexity vanishes towards and close to the crunch, but the time dependence is no longer linear in this case. The considerations to be applied  in constraining the proximity of the time to the singularity while retaining the validity of the approximation are similar to those
presented for the Kasner metrics.

We will now spend a moment on an issue of the ${\cal C} \propto  {\cal V}$ not discussed previously. 
When we have only one boundary for the CFT frame, a quantum state is defined once we specify one number, the time $t^*$ when we take the spatial slice. 
The maximal volume equation (\ref{eqf}) on the other hand is a second order differential equation and needs two conditions to select a particular volume that extends in the bulk.  This second boundary condition, for $d>1$,  appears in the expansion (\ref{expf}) to higher order than ${\cal O}(z^2)$, so it does not affect either the leading (\ref{vol}) or the sub-leading (\ref{volsub}) divergences  of the volume. Nevertheless, if we want to make sense of the  ${\cal C} \propto  {\cal V}$ conjecture we need to specify this condition and select one particular volume. 
If we take $h_{ij}$ constant, the generic solution of (\ref{eqf}) with $f(0) = t^*$ is
\beq
f(z) = t^* + z \psi(\beta z^{d+1}) =   t^* + \frac{1}{d+2} \beta z^{d+2}  + \dots
\eeq 
where $\psi$ is some known function expressed in terms of an hypergeometric function and $\beta$ is the second integration constant which appears only at order $z^{d+2}$ in the expansion.  
All of those solutions are locally maximal volume surfaces  and the union of all of them gives the WdW patch.
If we want to select a particular one, in order to associate it to the complexity of the boundary state, a natural choice would be the constant solution, $\beta = 0$,  equal to $t^*$ everywhere. 
\begin{figure}[h]
\begin{center}
\includegraphics[scale=.5]{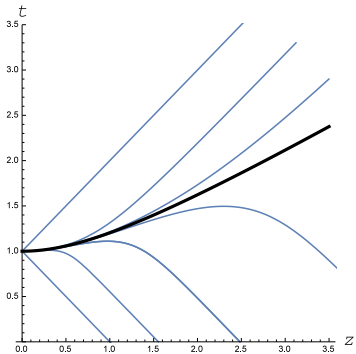} 
\end{center}
\caption{\small  Solutions to the maximal volume equation (\ref{eqf}) for the case of the Kasner metric for  $d=2$ and $t^*=1$.   }
\label{maxsurf}
\end{figure}
When $h_{ij}$ depends of time this choice becomes less obvious. Let's take for example the case of Kasner and we plot in Fig. \ref{maxsurf} various solutions of the maximal volume equation 
(\ref{eqf}) with $f(0) = t^*$. As before the union of all the solutions gives the WdW patch.  All solutions, save one, become asymptotically null at large $z$ like $f(z) =   z + {\rm const}$ or $f(z) =- z + {\rm const}$.  The special one that divides the two types of solutions is a natural candidate to be associated at the boundary state.  Its slope at $z \to \infty$ in neither $1$ nor $-1$ and in general depends on the anchored time $t^*$. 
For the purpose of the present paper this is not an urgent issue since all those solutions share the same quadratic expansion and thus have the same leading and sub-leading divergences. So we will  not discuss this issue any further here.

\begin{figure}[h]
\centering
\includegraphics[scale=.4]{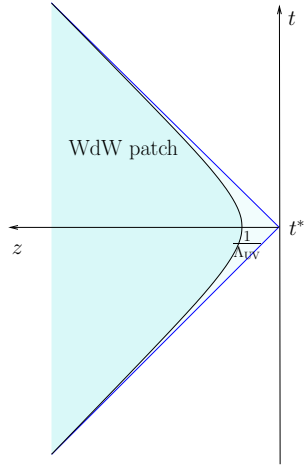}
\caption{\small    Regularization of the WdW patch. }
\label{figuno}
\end{figure}

We now turn to obtain the complexity as given by the definition ${\cal C} \propto  {\cal A}$  where ${\cal A}$  is the action of the WdW patch anchored at $t = t^*$ on the UV boundary. 
We need to regularize the action and to this end we smoothen-out the boundary of the patch. Our choice is to take,  as in Figure \ref{figuno}, the region of space-time delimited by the following  hyperbole 
\beq
\label{bou}
z^2 - (t-t^*)^2 = \frac{1}{\Lambda_{\rm UV}^2}  \ .
\eeq
As $\Lambda_ {\rm UV} \to \infty$ we recover the full WdW patch. 
This regularization achieves two goals simultaneously. First it regularize the UV divergence. Second the boundary of the regularized patch is now a smooth time-like boundary with no  joints: this allows the computation of the boundary gravitation action without any ambiguities. 

The total action is given by the Einstein-Hilbert (EH) term with the cosmological constant plus the boundary York-Gibbons-Hawking (YGH) term
\bea
{\cal A} &=& 
{\cal A}_{EH} + 
{\cal A}_{YGH} \nonumber \\
 &=& \frac{1}{16 \pi G} \int_{WdW}  \sqrt{ - g}\,( R - 2 \Lambda) + \frac{\epsilon}{8 \pi G} \int_{\partial WdW}  \sqrt{|\gamma |}\, K 
\eea
where $\epsilon =\pm 1$ according if the boundary is time-like or space-like. The boundary of the WdW is a null hypersurface, but we will consider it as a limit of a time-like submanifold (\ref{bou}). So from now on we will assume $\epsilon = + 1$ and $|\gamma| = -\gamma$.

The EH term evaluated on the regularized patch  is  
\bea
\label{ehuno}
{\cal A}_{EH} &=&  \frac{1}{16 \pi G} \int dt  \int_{ \frac{\sqrt{ 1 +{\Lambda_{\rm UV}^2 (t-t^*)^2 }}}{\Lambda_{\rm UV}}
}^{\infty} dz   \int dV_x \, \frac{L^{d+2}}{z^{d+2}}  \sqrt{  h(t,x)} ( R - 2 \Lambda) \nonumber \\
&=&  \frac{1}{16 \pi G} \int dt     \int dV_x \, \frac{L^{d+2} \Lambda_{\rm UV}^{d+1}}{(d+1) ( 1 +{\Lambda_{\rm UV}^2 (t-t^*)^2 })^{\frac{d+1}{2}}}  \sqrt{  h(t,x)} ( R - 2 \Lambda) 
\eea
We are interested in capturing  the dominant divergent piece of the action.
So we take $g$ to be equal to its boundary value at $t=t^*$:
\bea
\label{ehdue}
{\cal A}_{EH} &\simeq&  -  \frac{1}{8 \pi G}  \frac{ \Lambda_{\rm UV}^{d} L^{d} \sqrt{\pi}\Gamma(\frac{d}{2})}{ \Gamma(\frac{d+1}{2})  } \int dV_x \, \sqrt{  h(t^*,x)}  \ .
\eea
Note that this term is proportional to the boundary space volume 
and the divergence is proportional to $\Lambda_{\rm UV}^{d}$. The  sign of ${\cal A}_{EH}$ is  negative because of the sign of  $R -2 \Lambda$ , but it will be overcome by the boundary term which turns out to be positive.
To see to the final answer the reader can  go directly to (\ref{yghdue}) and (\ref{finalaction}). In the following we present the technical details needed for obtaining ${\cal A}_{YGH}$.

We need to compute some useful geometrical quantities before the evaluation of the boundary YGH action.
The  components of the connection we will  use later are:
\bea
\Gamma^z_{zz}  = - \frac{1}{z}\ , \qquad \Gamma^t_{zt} = \Gamma^t_{tz}  = \Gamma^z_{tt} =- \frac{1}{z}\ , \qquad \Gamma^z_{ij} = \frac{1}{z}h_{ij} \ ,  \qquad \Gamma^t_{ij}  =   \frac{1}{2} \partial_th_{ij} \ .
\eea
The boundary (\ref{bou}) is a time-like sub-manifold and we can choose to parameterize it with the coordinates $x^a: t,x_i $. 
It is defined by the following embedding:
\beq
\label{hyperboloid}
x^a: \left(t,x_i\right) \quad \longrightarrow \quad x^{\mu} : \left(t,z= \frac{1}{\Lambda_{\rm UV}}\sqrt{ 1 +{\Lambda_{\rm UV}^2 (t-t^*)^2 }},x_i\right) \ .
\eeq
The Jacobian of the embedding $e^{\mu}_a = \frac{\partial x^{\mu}}{\partial x^a}$ is
\bea
\label{jacobian}
&&e^{t}_{t} = 1 \quad e^{z}_{t} = \frac{\Lambda_{\rm UV}(t-t^*)}{\sqrt{ 1 +\Lambda_{\rm UV}^2 (t-t^*)^2 }} \quad e^{i}_{t} = 0  \nonumber \\
&&e^{t}_{j} = 0 \quad e^{z}_{j} = 0 \quad  e^{i}_{j} = \delta^i_j  \ .
\eea
The induced metric $\gamma_{ab}$ on the boundary manifold $x^a$ is 
\beq
ds^2 = \frac{L^{2}}{z^2} \left(- \frac{1}{(t-t^*)^2\Lambda_{\rm UV}^2+ 1} dt^2  + h_{ij}(t,x) dx_i dx_j  \right) \ .
\eeq
The  outward unit  normal is
\bea 
 && n_t =  \frac{L \Lambda_{\rm UV}^2(t-t^*)}{\sqrt{1 +\Lambda_{\rm UV}^2 (t-t^*)^2 }} \qquad \qquad \ 
n_z = -L \Lambda_{\rm UV} 
\eea
with $e^{\mu}_a n_{\mu} = 0$ and $ n^{\mu}n_{\mu} = 1$. 
The final result will not depend on the specific way $n^{\mu}$ is extended outside the boundary.
The covariant derivative is $ \nabla_{\mu} n_{\nu} = \partial_{\mu}  n_{\nu} - \Gamma^{\rho}_{\mu \nu} n_{\rho} $,  and the components are
\bea
& &\nabla_{t} n_{t} = \partial_t n_t +\frac{n_z}{z} \  \qquad 
 \nabla_{z} n_{t} =  \nabla_{t} n_{z} =  \frac{n^t}{z} \ \nonumber \\
 & &\nabla_{z} n_{z} = \frac{n_z}{z} \ \qquad \qquad \ \ \, 
\nabla_{i} n_{j} =- \frac{1}{2  } \partial_t h_{ij }n_t - \frac{1}{z }  h_{ij }n_z  \ .
\eea
We can now compute the extrinsic scalar curvature of the boundary sub-manifold
\bea
K &=& \gamma^{ab} e^{\mu}_a e^{\nu}_b  \nabla_{\mu} n_{\nu} \nonumber \\
 &=&   \frac{1}{L} \left( d - \frac12 h^{ij} \partial_t h_{ij}(t-t^*) \right)\sqrt{1+(t-t^*)^2\Lambda_{\rm UV}^2 } \ .
\eea
The YGH boundary term is then
\bea
\label{yghuno}
{\cal A}_{YGH} &=& \frac{1}{8 \pi G} \int dt   \int dV_x \,  \frac{L^{d+1}\Lambda_{\rm UV}^{d+1}}{((t-t^*)^2\Lambda_{\rm UV}^2+ 1)^{\frac d2 +1}}   \sqrt{  h(t,x)} K \ .
\eea
As before, to extract the dominant divergent term  we take $g$ to be equal to its  boundary value at $t=t^*$:
\bea
\label{yghdue}
{\cal A}_{YGH} &\simeq& \frac{1}{8 \pi G} \int ds \frac{L^{d}\Lambda_{\rm UV}^{d+1} d }{(s^2\Lambda_{\rm UV}^2+ 1)^{\frac{d+1}{2}}}   \int dV_x \, \sqrt{  h(t^*,x)}    \nonumber \\ &\simeq&  \frac{1}{8 \pi G}
 \frac{L^{d}\Lambda_{\rm UV}^d  d \sqrt{\pi}\Gamma(\frac{d}{2})}{\Gamma(\frac{d+1}{2})} \int dV_x \, \sqrt{  h(t^*,x)} \ .
\eea

The final result for the dominant divergent term of the action  is then the sum of (\ref{ehdue}) and (\ref{yghdue}):
\beq
\label{finalaction}
{\cal A} \simeq \frac{L^{d} \Lambda_{\rm UV}^d}{8 \pi G} \frac{  \sqrt{\pi} \Gamma(\frac{d}{2}) (d-1) }{\Gamma(\frac{d+1}{2})}  \int dV_x \, \sqrt{  h(t^*,x)} \ .
\eeq
The coefficient is always positive for $d > 1$ despite the EH being negative, the positive  YGH term is always the dominant contribution.
The result for the leading UV divergence of the complexity is thus the same obtained  for the ${\cal C}  \propto {\cal V}$  in  (\ref{vol}). In both cases the complexity decreases towards the singularity in the region of validity of the
approximation.

We next study the sub-leading divergences. 
For this we need the expansion of the metric determinant up to the second order around $t^*$
\beq
h(t,x) \simeq  h(t^*,x) \left( 1+  (t-t^*) H_1(t) +  \frac{1}{2}  (t-t^*)^2   H_2(t) + \dots \right)
\eeq
where $H_1(t)$ and $H_2(t)$ are given by
\beq
 H_1(t) =  h^{ij}\partial_t h_{ij} \ , \qquad  H_2(t) = h^{ij}\partial^2_t h_{ij} -   h^{jl}\partial_t h_{lk} h^{ki}\partial_t h_{ij}+ H_1^2 \ .
\eeq
The  expansion for the EH terms (\ref{ehuno}) gives  
\bea
\label{td}
{\cal A}_{EH} \simeq {\cal O}( \Lambda_{\rm UV}^d)   -  \frac{L^{d} \Lambda_{\rm UV}^{d-2}}{128  \pi G}  \frac{ \sqrt{\pi}\Gamma(\frac{d}{2}-1)}{ \Gamma(\frac{d+1}{2})  } \left(2 H_2 -H_1^2 \right) \int dV_x \, \sqrt{  h(t^*,x)}
\eea
where ${\cal O}( \Lambda_{\rm UV}^d)$ is the leading term (\ref{ehdue}). 
The  expansion for the YGH terms (\ref{yghuno}) gives  
\bea
\label{td}
{\cal A}_{YGH} \simeq {\cal O}( \Lambda_{\rm UV}^d)  +\frac{L^{d} \Lambda_{\rm UV}^{d-2}}{128 \pi G}   \frac{ \sqrt{\pi}\Gamma(\frac d2-1)}{ \Gamma(\frac{d+1}{2})} \left(2 d  H_2 - (d+2) H_1^2 \right) \int dV_x  \sqrt{  {\rm det}\,  h_{ij}(t^*,x)} 
\eea
where ${\cal O}( \Lambda_{\rm UV}^d)$ is the leading term (\ref{yghdue}). 
The  final result is 
\bea
\label{td}
{\cal A}\simeq {\cal O}( \Lambda_{\rm UV}^d)  +\frac{L^{d} \Lambda_{\rm UV}^{d-2}}{128 \pi G}   \frac{ \sqrt{\pi}\Gamma(\frac d2-1)}{ \Gamma(\frac{d+1}{2})} \left(2 (d-1)  H_2 - (d+1) H_1^2 \right) \int dV_x  \sqrt{  {\rm det}\,  h_{ij}(t^*,x)} \nonumber \\
\eea
where ${\cal O}( \Lambda_{\rm UV}^d)$ is the leading term (\ref{finalaction}).


For the  Kasner metric $H_1 = \frac{2}{t} $, $H_2 = \frac{2}{t^2} $ and the action is then
\beq
{\cal A} \simeq \frac{L^{d} \Lambda_{\rm UV}^d  V t }{8\pi G l}  \frac{  \sqrt{\pi} \Gamma(\frac{d}{2}) (d-1 ) }{\Gamma(\frac{d+1}{2})} -\frac{L^{d} \Lambda_{\rm UV}^{d-2} V }{16 \pi G l t}   \frac{ \sqrt{\pi}\Gamma(\frac d2-1)}{ \Gamma(\frac{d+1}{2})}  \ .
\eeq
In this case, due to the fact that $\sqrt{h}$ is linear in $t$, we essentially recover the same result we had for the volume prescription (\ref{volsubkas}).
For the  case of topological crunch instead we see some difference in the subleading divergence because $\sqrt{h}$ is not linear in time. In the latter case we have $H_1 =-\frac{2}{l} \tan{\left( {\tiny \frac tl} \right)}$, $H_2 =\frac{2}{l^2} \left( \tan^2{\left( {\tiny \frac tl} \right)} -1 \right)$ and the action is 
\bea
{\cal A} &\simeq& \frac{L^{d} \Lambda_{\rm UV}^d  l^d V_{S^{d}}  \cos{\left( {\tiny \frac tl} \right)}  }{4 G} \frac{  \sqrt{\pi} \Gamma(\frac{d}{2}) (d-1) }{\Gamma(\frac{d+1}{2})}  + \nonumber \\
&& \qquad -  \frac{L^{d} \Lambda_{\rm UV}^{d-2}   l^{d-2} V_{S^{d}}  \cos{\left( {\tiny \frac tl} \right)}  \left(2   \tan^2{\left( {\tiny \frac tl} \right)} + d-1 \right) }{16  G}   \frac{ \sqrt{\pi}\Gamma(\frac d2-1)}{ \Gamma(\frac{d+1}{2})} \ .
\eea
We see that the subleading divergence is different from the one obtained with the volume prescription (\ref{volsubtop}).

In the next part of the paper we will  test if the result we have found in this section, that the main features of the complexity for space like singularities, are independent on the choice among the two bulk prescriptions, is valid also for the de Sitter case.

\section{Time dependence in the bulk:  De Sitter crunch}
\label{tre}

We next turn to cases in which  a Big Crunch occurs semi-classically in the bulk.  This can be represented by either a boundary theory, with particular time dependent relevant deformations, living on a time independent  world volume which exists for only a finite time (denoted EF) or by a conformally related time independent boundary system living on an expanding de Sitter world volume (denoted by dSF). This duality was studied in \cite{Barbon:2010gn, BR}.  In this case the time derivative of the complexity is a finite quantity, both for the volume and the action prescriptions. It does not depend on the UV regularization.

We first make a more generic analysis in the absence of divergent terms in the time derivative of the complexity. 
Consider a  metric is of the form  
\beq
\label{gm}
ds^2 = \frac{L^2}{z^2} \left( dz^2 + h_{\mu \nu}(t,x) dx^{\mu} dx^{\nu}  \right) 
\eeq
with $\mu,\nu= 0,1, \dots, d$  and the UV boundary is at $z=0$. The expansion of the metric near the boundary is in general of the following form
\beq
 h_{\mu \nu} = h^{(0)}_{\mu\nu} + z^2 h^{(2)}_{\mu\nu} + \dots + z^d h^{(d)}_{\mu\nu} + z^d \log{z^2} h^{(d)}_{\mu\nu} +\dots 
\label{expansionm}
\eeq
For the EF case one has   $h^{(0)}_{\mu\nu}$ equal to the EF metric and the entire time dependence is due to a domain wall moving in the bulk.  We thus consider the case in which  $h^{(0)}_{\mu\nu}$  is static while  sub-leading terms in the expansion (\ref{expansionm}) may contain information about the time dependence.
Let's assume for generality that $h^{(2k)}_{\mu\nu}(t,x)$ is the first term to contain a time dependence, then the  contribution to the complexity from the term (\ref{td}) is
\bea
{\cal A} &\simeq& {\cal O}( \Lambda_{\rm UV}^d) +  {\cal O}( \Lambda_{\rm UV}^{d-4k}) (\partial_t h^{(2k)})^2 
\eea
So, for small enough dimension $d<5$
we do not have any divergent terms.\footnote{Rember that $d$ is the number of space dimensions on the boundary, for example $d=3$ for AdS$_5$. For $d \geq 5$, so from AdS$_7$ onward, a bulk time dependence of the metric can in general provide a time-dependent divergent piece in the action trough its tail at $z \to 0$. We will not consider those cases in the present paper. }
In this case to compute the time dependence of the complexity, which is a  finite term,  we have  to consider the full metric and also the precise shape of the WdW patch; considering only the asymptotic form of the metric close to the UV is not enough.
This leads us to perform the computations on a case-by-case basis.

We now consider in detail the de Sitter crunch in the case that the thin wall approximation is valid. This occurs  when the boundary field theory in its dSF coordinates contains 
a negative relevant operator whose coefficient is much larger than its expansion rate \cite{Barbon:2010gn,BR}.
The metric in the UV is AdS$_+$ with curvature length $l_+=1$ which we fix to one for convenience
\beq
\label{mro}
ds^2 = d\rho^2 + \sinh^2{(\rho)} \left ( -d \tau^2 + \cosh^2{(\tau)} d\Omega_{d}^2 \right) 
\eeq
This is valid for $\rho > \rho_W$ where $\rho_W$ is the position of the domain wall. 
The metric on the other side of the wall  $\rho < \rho_W$   is AdS$_-$ with a smaller curvature length $l_- < 1$. 
In order to match the two metrics at the same value of $\rho$ we need to perform some rescaling of the standard form. The expression for the interior metric is
\beq
\label{intrho}
ds^2 =  l_-^2 \alpha^2  d\rho^2 + l_-^2  \sinh^2{(\alpha \rho)} \left ( -d \tau^2 + \cosh^2{(\tau)} d\Omega_{d}^2 \right)
\eeq
with
\beq
\alpha =  \frac{1}{\rho_W} {\rm arcsinh} \left(  \frac{ \sinh{( \rho_W)}}{l_-}  \right) > 1
\eeq
This is valid for $\rho < \rho_W$. In this coordinate choice the two metrics are patched so that they are continuous at $\rho = \rho_W$. The position of the wall $\rho_W$ is constant in the time $\tau$. In the dual field theory this correspond to an RG flow between two different CFT's.  The limit we are discussing is that of the thin-wall approximation.


Now we focus on  the exterior region, the one that extends to the UV.
By changing coordinates to
\beq
r = \sinh{(\rho)} \cosh{(\tau)} \qquad \cos{( t )} = \frac{ \cosh{(\rho)}}{\sqrt{1 +\sinh^2{(\rho)} \cosh^2{(\tau)} }}
\eeq
the metric (\ref{mro}) becomes
\beq
\label{metext}
ds^2 = \frac{dr^2}{\left(1+ r^2 \right)} -\left(1+r^2\right) dt^2 + r^2 d\Omega_{d}^2
\eeq
and the wall is not static but follows the trajectory
\beq
\label{wallext}
r(t)_{W} =  \left(\frac{\cosh^2{(\rho_W)}}{\cos^2{(t)}}-1 \right)^{\frac12}
\eeq
At the time $t \to \frac{\pi}{2}$ where $r(t)_W \to \infty$  
we have a  bulk crunch. We will work in this EF frame where the metric does not depend on time, although it exists only for a finite time,  and the bulk crunch is visible on the boundary. Note that the metric time dependence is only through the wall trajectory. In the UV the metric is constant.

We then have consider the interior part AdS$_-$ with  $L=l_-$. We first  rewrite (\ref{intrho}) the metric as 
\beq
ds^2 = l_-^2 \left( d\trho^2 +  \sinh^2{(\trho)} \left ( -d \tau^2 + \cosh^2{(\tau)} d\Omega_{d}^2 \right) \right)
\eeq
where
\beq
\trho = \alpha \rho
\eeq
The domain wall is then at $\trho_W = \alpha \rho_W$.
Then we change coordinates to
\beq
\tr = \sinh{(\trho)} \cosh{(\tau)} \qquad \cos{\left(\tt \right)} = \frac{ \cosh{(\trho)}}{\sqrt{1 +\sinh^2{(\trho)} \cosh^2{(\tau)} }}
\eeq
and the metric becomes 
\beq
\label{metint}
ds^2 = l_-^2 \left(\frac{d\tr^2}{1+ \tr^2 } -\left(1+\tr^2\right) d\tt^2 + \tr^2 d\Omega_{d}^2 \right)
\eeq
The wall trajectory is these coordinates is
\beq
\label{wallint}
\tr(\tt)_{W} =  \left(\frac{\cosh^2{(\alpha \rho_W)}}{\cos^2{\left(\tt \right)}}-1 \right)^{\frac12}  
\eeq
The metric is  AdS$_-$ in the region  $\tr \leq \tr(\tt)_W$. The crunch is at $\tt  \to   \frac{\pi}{2}$ where $\tr(\tt)_W \to \infty$.

In what follows we will use two different coordinate systems: for the exterior (\ref{metext}) and for the interior (\ref{metint}). The two patches are connected at the wall trajectory which is different in the two coordinates systems, respectively (\ref{wallext}) and (\ref{wallint}). This is the price we pay in order to have the two metrics writen in a simple form. When passing from one patch to the other we have to change also coordinates. A point on the domain wall in AdS$_+$ at time $t$ and radius $r(t)_W$ in the AdS$_-$ patch will be seen at time and radius
\beq
\label{rel}
\tt = \arccos{\left( \frac{\cos{(t)} \cosh{(\alpha \rho_W)}}{
\sqrt{\cos^2{(t)}+\frac{\sinh{(\alpha \rho_W)}}{\sinh{( \rho_W)}}\left(\cosh^2{\left(\rho_W\right)}-\cos^2{(t)}\right)
}}\right) }\ ,\qquad \tr = \frac{r}{l_-} \ .
\eeq

 We work in the thin wall approximation.  In general we can construct a wall with a scalar field and a suitable potential with two minima, the values at the two minima correspond to the cosmological constant at the two sides of the wall. 
We can then scale the microscopic parameters and send the wall thickness to zero while keeping the tension fixed. Assuming that the curvature in the middle of the wall remains of the same order of the cosmological constants, the thin wall limit allows us to neglect the terms from the action  coming from the wall itself.

The space volume anchored at $t=t^*$ in the UV part is
\bea
{\cal V_+} &=& \int_{r_{W}(t^*)}^{\Lambda_{\rm UV}}  dr  \frac{r^{d}}{\sqrt{1+r^2}} V_{S^{d}}  
\eea
The space volume anchored at $t^*(t^*)$ (the relation is given by (\ref{rel})) in the IR part is
\bea
{\cal V_-} = \int_0^{\tr_{W}(t^*)}  dr   \frac{l_-^{d+1} r^{d}}{\sqrt{1+r^2 }} V_{S^{d}} 
\eea
The total volume is then 
\beq
{\cal V} = {\cal V_+} + {\cal V_-} = V_{S^{d}}  \left( l_-^{d+1} \int_0^{\tr_{W}(t^*)  }  dr   \frac{r^{d}}{\sqrt{1+r^2 }}+
 \int_{r_{W}(t^*)}^{\Lambda_{\rm UV}}  dr  \frac{r^{d}}{\sqrt{1+r^2}}    \right)
\eeq
The derivative with respect to $t^*$ is free from UV divergence and, using (\ref{rel}), it becomes
\beq
\frac{ d{\cal V}}{dt^*} =  V_{S^{d}}  \frac{dr_{W} }{d t}  r_{W}(t^*)^{d} \left(
 \frac{1}{\sqrt{1+r_{W}(t^*)^2/l_-^2}} - 
 \frac{1}{\sqrt{1+r_{W}(t^*)^2}}  \right)
\eeq
This is always negative for $l_-<1$ and $t > 0$ which means that complexity(volume) is decreasing going toward the crunch. 
For $t^* \to \frac{\pi}{2}$ we have
\beq
\label{volclose}
\frac{ d{\cal V}}{dt^*} =  V_{S^{d}}  \frac{\cosh^{d}{(\rho_w)} (-1+l_-) }{\left(\frac{\pi}{2} - t^*\right)^{d+1}} 
\eeq
which is negative and becomes larger and larger close to the crunch as $\propto \frac{1}{\left(\frac{\pi}{2} - t^*\right)^{d+1}} $.

\begin{figure}[h]
\centering
\includegraphics[scale=.5]{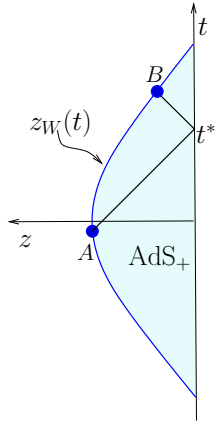}
\caption{\small The region AdS$_+$ enclosed between the UV boundary and the wall trajectory. We also show the WdW patch anchored at $t^*$ enclosed in this region and the intersections $z_A(t^*)$ and $z_B(t^*)$.  }
\label{figdue}
\end{figure}

We now  calculate the Complexity using the WdW prescription.
For this task it is convenient to change again coordinates
\beq
\label{changerz}
r =  {\rm cot}{(z)}
\eeq
with $0\leq z \leq  \frac{\pi}{2}$.
The metric  (\ref{metext}) becomes
\bea
\label{zmet}
ds^2 = \left(1+ {\rm cot}^2{(z)} \right) \left(  -  dt^2 + dz^2 + \frac{ {\rm cot}^2{(z)}}{1+ {\rm cot}^2{(z)}}  d\Omega_{d}^2 \right)  
\eea
The domain wall trajectory is 
\beq
z_{W}(t) =  \frac{\pi}{2} -\arctan{r(t)_{W}}
\eeq
Close to the crunch
\beq
z_{W}(t) \simeq \frac{\frac{\pi}{2} - t}{\cosh{(\rho_W)}} 
\eeq
In these coordinates the WdW patch is  delimited by 
\beq
\label{patchzt}
z= |t - t^*|
\eeq
Intersection with the domain wall trajectory is at $(t_A(t^*),z_A(t^*))$ and $(t_B(t^*),z_B(t^*))$ with $t_B(t^*) > t_A(t^*)$ and 
\beq
z_A(t^*) > z_B(t^*)  \qquad {\rm for} \qquad t^* >0 
\eeq
In general we cannot solve analytically for $z_A(t^*)$ and $z_B(t^*)$.
Close to the crunch the two intersections are
\bea&&t_A(t^*) \simeq   \frac{ \cosh{(\rho_W)} t^* - \frac{\pi}{2}}{ \cosh{(\rho_W)}-1}  \qquad \qquad \quad \ \ \  z_A(t^*) \simeq    \frac{\frac{\pi}{2} -t^*}{\cosh{(\rho_W)}-1} 
 \nonumber \\
&&t_B(t^*) \simeq   \frac{  \cosh{(\rho_W)} t^* + \frac{\pi}{2}}{\cosh{(\rho_W)}+1} \qquad \qquad \quad \ \ \  z_B(t^*) \simeq    \frac{\frac{\pi}{2} -t^*}{ \cosh{(\rho_W)} +1}  
\eea
The geometry of AdS$_+$ with the WdW patch is pictured in Figure \ref{figdue}.

Now we compute the action of the WdW patch anchored at the UV boundary $t^*$. 
The EH term is  
\bea
{\cal A}_{EH_+}  &=&  \frac{1}{16 \pi G} \int_{WdW}  \sqrt{ -g} ( R - 2 \Lambda)
\nonumber \\&=& -  \frac{(d+1) V_{S^{d}}}{8\pi G}  \int_{t_A}^{t_B} dt \int_{|t-t^*|}^{z_W(t)}  dz \,  {\rm cot}^{d}{(z)}} (1+ {\rm cot}^2{(z)) \nonumber \\
\eea
where  we used $R - 2 \Lambda = -2 (d+1)$. The derivative with respect to $t^*$ is
\bea
\label{daehp}
\frac{d{\cal A}_{EH_+}}{dt^*}  &=&  \frac{ V_{S^{d}}}{8 \pi G  }   
\left(  {\rm cot}^{d+1}{(z_B)} -  {\rm cot}^{d+1}{(z_A)} \right) 
\eea

The boundary  YGH term is obtained next . The metric (\ref{zmet}) is
\bea
ds^2 = g(z) \left(  -  dt^2  + dz^2\right) + f(z) h_{ij}(x) dx^i dx^j   
\eea
with
\beq
g(z) = 1+ {\rm cot}^2{(z)}  \qquad f(z) =  {\rm cot}^2{(z)}
\eeq
and $h_{ij}$ the metric of $S^{d}$.
The  component of the connection of the metric that we need are:
\bea
\Gamma^z_{zz}  = \frac{g'(z) }{2 g(z)}\  \qquad \Gamma^t_{zt} = \Gamma^t_{tz} =\Gamma^z_{tt}  = \frac{g'(z)}{2 g(z)}\qquad 
 \Gamma^z_{ij} =  -\frac{ f'(z) }{2g(z)}h_{ij} \ .
\eea
We use the same time-like hyperboloid  as before and then send the cutoff to infinity to obtain the light-like boundary of WdW.
The boundary  is defined by (\ref{hyperboloid}).
The induced metric $\gamma_{ab}$ on $x^a$ is 
\beq
ds^2 = - \frac{g(z)}{(t-t^*)^2\Lambda_{\rm UV}^2+ 1} dt^2  + f(z)  h_{ij}(t,x) dx_i dx_j  
\eeq
The Jacobian of the embedding $e^{\mu}_a =\frac{ \partial x^{\mu}}{\partial x^a}$ is the same as in (\ref{jacobian}).
The  outward unit  normal is
\bea
n_t =  \sqrt{g(z)} \Lambda_{\rm UV}(t-t^*) \qquad
n_z = - \sqrt{g(z) (1 +\Lambda_{\rm UV}^2 (t-t^*)^2)} \ .
\eea
The  components of the  covariant derivative
 different from zero are
\bea
 && \nabla_{t} n_{t} = \partial_t n_t - \frac{g'(z)}{2 g(z)} n_z \qquad 
 \nabla_{z} n_{t} =  \partial_z n_t  - \frac{g'(z)}{2 g(z)} n_t \nonumber \\
 && \nabla_{t} n_{z}  =  \partial_t n_z   - \frac{g'(z)}{2 g(z)} n_t \qquad 
 \nabla_{z} n_{z} = \partial_z n_z - \frac{g'(z)}{2 g(z)} n_z  \nonumber \\
 && \qquad \qquad  \qquad  \quad \nabla_{i} n_{j} =  \frac{ f'(z) }{2g(z)}h_{ij} n_z 
\eea
and the extrinsic curvature is then 
\beq
K = \gamma^{ab} e^{\mu}_a e^{\nu}_b  \nabla_{\mu} n_{\nu} =- \frac{\Lambda_{\rm UV}}{g(z)^{\frac12}} - \frac{d z\Lambda_{\rm UV} f'(z)}{2 f(z) g(z)^{\frac12}} - \frac{\Lambda_{\rm UV} z g'(z)}{2 g(z)^{\frac32}}  \ .
\eeq 
Putting all these together allows to be obtain the YGH boundary term:
\bea
{\cal A}_{YGH+} &=& \frac{1}{8 \pi G} \int_{\partial WdW}  \sqrt{ - \gamma} K
\nonumber \\&=& \frac{ V_{S^{d}}}{8 \pi G} \int dt   \frac{g(z)^{\frac12} f(z)^{\frac{d}{2}}}{((t-t^*)^2\Lambda_{\rm UV}^2+ 1)^{\frac12}}    K 
\eea
The terms is simplified  by  removing the cutoff, that is sending $\Lambda_{\rm UV} \to \infty$ 
\bea
{\cal A}_{YGH+}  &=&  - \frac{ V_{S^{d}}}{16\pi G} \int_{t_A}^{t_B} dt  f(z)^{\frac{d-2}{2}} \left(  \frac{2f(z)}{z} + d f'(z) + \frac{f(z) g'(z)}{g(z)}\right)
\eea
where $z = |t - t^*|$. 
The time derivative takes contributions only from the boundary of the boundary since the bulk of AdS$_+$ in time independent  and is 
\bea
\label{dayghp}
\frac{d{\cal A}_{YGH+}}{dt^*}  &=& -  \frac{ V_{S^{d}}}{16 \pi G} \left(  \frac{dz_B}{dt^*} f(z_B)^{\frac{d-2}{2}} \left(   \frac{2 f(z_B)}{z_B} + d f'(z_B)+ \frac{f(z_B) g'(z_B)}{g(z_B)}\right) \right.  \nonumber \\&& \qquad  \quad \  \left.+ \frac{dz_A}{dt^*}   f(z_A)^{\frac{d-2}{2}} \left(   \frac{2f(z_A)}{z_A} + d f'(z_A) + \frac{f(z_A) g'(z_A)}{g(z_A)}\right)  \right) \ .
\eea

Next we evaluate  the WdW action in AdS$_-$ interior.
It is convenient to change coordinates as 
\beq
\label{trtz}
\tr = \cot{(\tz)}
\eeq
 and the metric (\ref{metint}) is then
\bea
\label{zmet}
ds^2 = l_-^2 \left(1+ {\rm cot}^2{(\tz)} \right) \left(  -  d\tt^2 + d\tz^2 + \frac{ {\rm cot}^2{(\tz)}}{1+ {\rm cot}^2{(\tz)}}  d\Omega_{d}^2 \right)  
\eea
The domain wall trajectory is 
\beq
\tz_{W}(\tt) = \frac{\pi}{2} -\arctan{\tr(\tt)_{W}}
\eeq
Intersection with the domain wall trajectory is at $\left(\tt_A,\tz_A\right)$ and $\left(\tt_B,\tz_B\right)$ to be obtained using (\ref{rel}) and (\ref{trtz}) from the corresponding values in AdS$_+$ which are $ (t_A(t^*),z_A(t^*))$ and $(t_B(t^*),z_B(t^*))$.
In these coordinates the WdW patch is again delimited by 
\beq
\tz=- \tt +\tt_A + \tz_A\qquad 
\tz= \tt -\tt_B + \tz_B
\eeq
together with $\tz_W(\tt)$.
The minimun and maximum values of $\tt$ are
\beq
\tt_{min} =\tt_A + \tz_A -\frac{\pi}{2}\qquad \tt_{max} = \tt_B - \tz_B + \frac{\pi}{2}
\eeq
The geometry of AdS$_-$ is pictured in Figure \ref{figtre}.
 \begin{figure}[h]
\centering
\includegraphics[scale=.4]{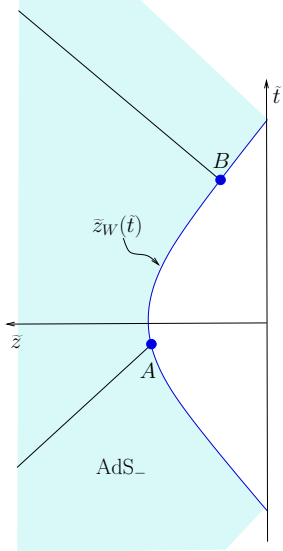}
\caption{\small   The region AdS$_-$ enclosed between the wall trajectory and the IR region. We also show the WdW patch.   }
\label{figtre}
\end{figure}

The action of the WdW patch that anchored at UV boundary $t^*$. 
The EH term is  
\bea
{\cal A}_{EH_-}  &=&  \frac{1}{16 \pi G} \int_{WdW}  \sqrt{ -g} ( R - 2 \Lambda)
\nonumber \\&=&  - \frac{(d+1) V_{S^{d}}  l_-^{d}}{ 8 \pi G}
\left(   \int_{\tt_{min}}^{\tt_A} d\tt \int_{- \tt +\tt_A + \tz_A}^{\pi/2}  d\tz  +   \int_{\tt_A}^{\tt_B} d\tt \int_{\tz_W(\tt)}^{\pi/2}  d\tz  +    \int_{\tt_B}^{\tt_{max}} d\tt \int_{\tt -\tt_B + \tz_B}^{ \pi/2}  d\tz \right)  \nonumber \\ 
&& \qquad \qquad \quad  
    {\rm cot}^{d}{\left( \tz \right)} \left(1+ {\rm cot}^2{\left( \tz \right)} \right) 
\eea
where  we used $R - 2 \Lambda = -\frac{2 (d+1)}{l_-^2}$. 
The derivative respect to $t^*$ is 
\bea
\label{daehm}
\frac{d{\cal A}_{EH_-}}{dt^*}   = -  \frac{ V_{S^{d}}l_- ^{d}}{8 \pi G} 
\left( \frac{d \left(\tt_B-\tz_B\right)}{dt^*}{\rm cot}^{d+1}{\left( \tz_B \right) } -  \frac{d \left(\tt_A+\tz_A\right)}{dt^*}{\rm cot}^{d+1}{\left( \tz_A \right) }  \right) 
\eea

The YGH boundary term is 
\bea
 {\cal A}_{YGH_-}  &=& - \frac{ V_{S^{d}}l_-^{d}}{16 \pi G}  \left( \int_{\tt_{min}}^{\tt_A}+  \int_{\tt_B}^{\tt_{max}} \right)   d\tt \nonumber \\
&& \qquad \qquad  f\left(\tz \right)^{\frac{d-2}{2}} \left(   \frac{2 f\left(\tz \right)}{\tz } + d f'\left(\tz \right) + \frac{f\left(\tz \right) g'\left(\tz \right)}{h'\left(\tz \right)}\right)
\eea
The time derivative is 
\bea
\label{dayghm}
\frac{d{\cal A}_{YGH_-}}{dt^*}  &=&    \frac{ V_{S^{d}}l_-^{d}}{16 \pi G} \left( \frac{d\tz_B}{dt^*} f\left(\tz_B \right)^{\frac{d-2}{2}} \left(   \frac{2f\left(\tz_B \right)}{\tz_B} + d f'\left(\tz_B \right) + \frac{f\left(\tz_B \right) g'\left(\tz_B \right)}{h'\left(\tz_B \right)}\right) \right.  \nonumber \\&& \qquad  \left.  +\frac{d\tz_A}{dt^*} f\left(\tz_A \right)^{\frac{d-2}{2}} \left(   \frac{2f\left(\tz_A \right)}{\tz_A} + d f'\left(\tz_A \right) + \frac{f\left(\tz_A \right) g'\left(\tz_A \right)}{h'\left(\tz_A \right)}\right)  \right) 
\eea

Finally we can now  sum all contributions to the derivative of the action
\beq
\frac{d{\cal A}}{dt^*} &=&
\frac{d{\cal A}_{EH}}{dt^*} +
\frac{d{\cal A}_{YGH}}{dt^*}
\eeq
which is the sum of the four terms (\ref{daehp}), (\ref{dayghp}), (\ref{daehm}), and (\ref{dayghm}).

We present a numerical evaluation of $\frac{d{\cal A}}{dt^*} $ in Figure \ref{figquattro}. The plots are done for the specific values of $d=2$,  $l_- = .9$ and $\rho_W = .5$. We also plot the  asymptotic formula close to the crunch (\ref{analiticala}).
 \begin{figure}[h]
\begin{center}
\includegraphics[scale=.8]{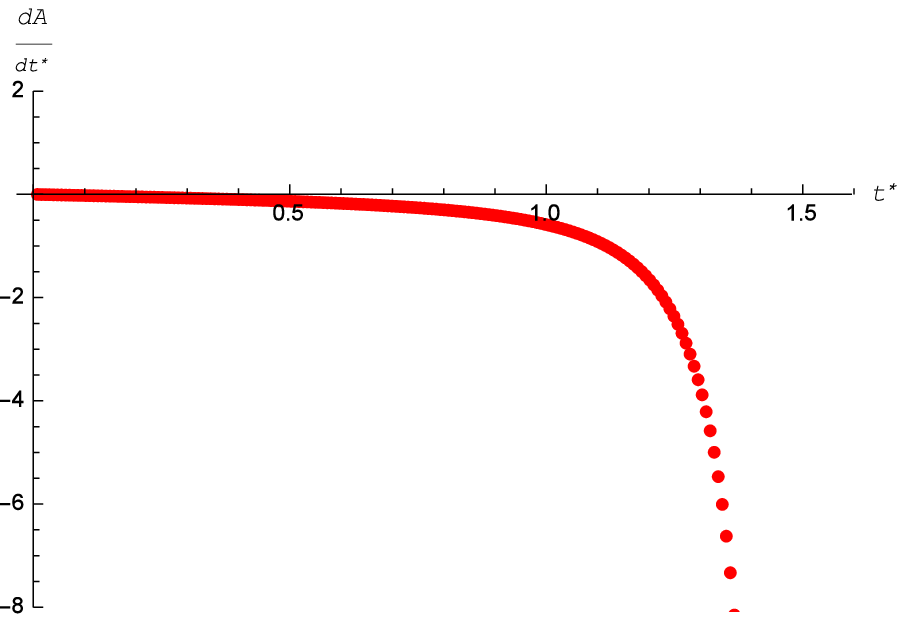} \quad
\includegraphics[scale=.8]{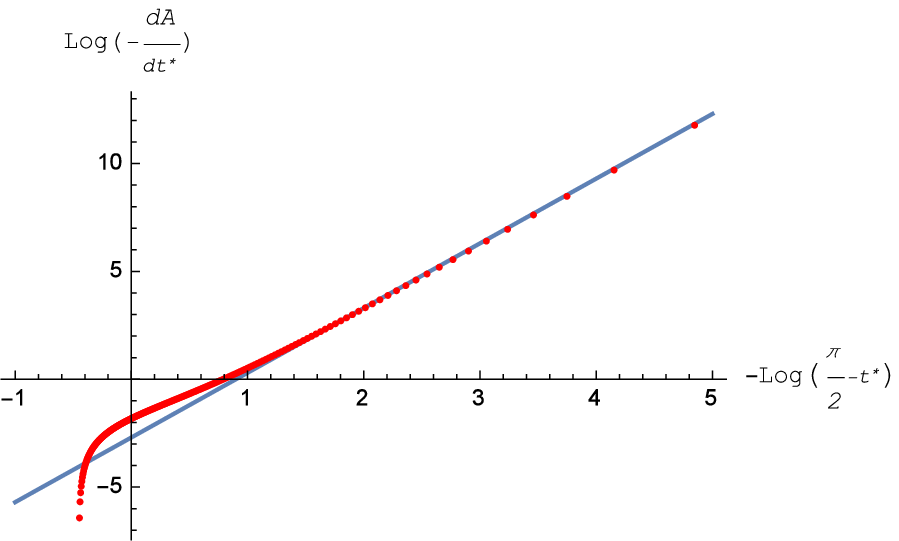} 
\end{center}
\caption{\small  $\frac{d{\cal A}}{dt^*} $  for the specific values  $d=2$, $l_- = .9$ and $\rho_W = .5$.  On the right, for the same parameters, the logarithmic plot of $\log{(-\frac{d{\cal A}}{dt^*})}$ vs. $-\log\left(\frac{\pi}{2} - t^*\right)$ which shows the validity of  (\ref{analiticala}).  }
\label{figquattro}
\end{figure}
\begin{figure}[h]
\begin{center}
\includegraphics[scale=.8]{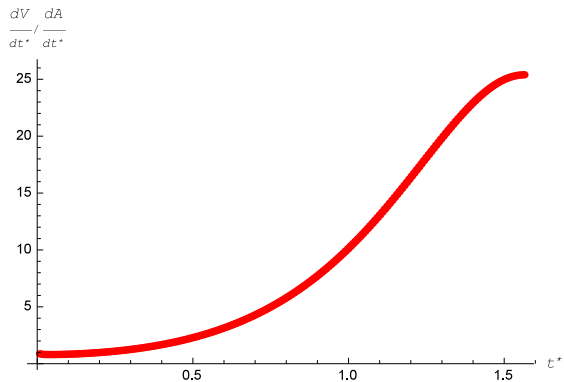} 
\end{center}
\caption{\small The ratio  $\frac{d{\cal V}}{dt^*} / \frac{d{\cal A}}{dt^*} $  for the same values of Figure \ref{figquattro}.   }
\label{figcinque}
\end{figure}
The time derivative is negative and becomes larger close to the singularity. 
That is qualitatively the same behaviour as obtained for the volume prescription (\ref{volclose})
\bea
\label{analiticala}
\frac{d{\cal A}}{dt^*} \propto \frac{1}{\left(\frac{\pi}{2} - t^*\right)^{d+1}}
\eea
The degree of approach to the singularity is not same as is shown in Figure \ref{figcinque} where we plot the ratio $\frac{d{\cal V}}{dt^*} / \frac{d{\cal A}}{dt^*}$. However the origin of the negative sign is the same in both cases, it follows because $l_-$, the curvature length in the IR,  is smaller than $l_+=1$ which is the curvature length at the UV. 
This  allows one to identify that the source of the decrease in complexity is the result of the renormalization group flow, there are less
degrees of freedom in the IR than in the UV.  Thus both definitions suggested for the Holographic complexity capture the same physical aspects
as far the decrease of complexity in these classes of singular cases.

\section{Conclusion}
\label{quattro}

Different motivations lie behind each of  the two suggestions for the bulk quantity which corresponds to the quantum holographic complexity of the boundary state. 
Given that  one does not have yet a precise and universal definition of complexity in quantum field theory itself, both can be considered as suggestive holographic probes for the notion of how complex is a quantum state.
 There is no a priori reason for them to coincide in a generic situation and indeed, while we have shown that the degree of divergence is the same, we have shown also that there  are quantitative differences. In some cases only
 for the next to leading term and in other cases for both the leading and next to leading term. The quantitative behaviour itself, even when the same
 for both suggestions, differed according the type of singular background. 
 However both deliver the same verdict as far as the manner in which the complexity of several classes of 
time-dependent holographic backgrounds, which have  crunch singularities, evolves. The complexity decreases. 
Another universal feature that we have found that the time derivative of the complexity  contains a UV divergent part if the boundary metric depends explicitly on time while it is finite otherwise. 
Another pattern  emerged from the action calculation. For the cases in which the time dependence character was driven mainly from a divergent term , it was
the boundary YGH term which determined the sign of the time derivative while for the case where the time dependence was driven by a finite
term in the evaluation of the complexity it was the EH volume term which dictated the final sign.
The universality we have found adds credence to the identification of the physical source of Complexity decrease given in \cite{Barbon:2015ria}.
This leaves us with yet one more indication that for some probes space like singularities may less of a bite than expected.




\section*{Acknowledgments}

We thank J. Barbon and J. Martin for useful discussions. We thanks Jie Ren for collaboration in the early stage of this project. S. B.  work
is supported by the INFN special research project grant GAST (``Gauge
and String Theories'').
 The work of E. R. was partially supported by the American-Israeli Bi-National Science Foundation.
 The work of  E. R.  and  S. R.  was partially supported by the  Israel Science Foundation Center of Excellence and the I Core Program of the Planning and Budgeting Committee and The Israel Science Foundation The Quantum Universe.
The work of S. R.  is  supported by the IIT Hyderabad seed grant SG/IITH/F171/2016-17/SG-47.

\end{document}